\pdfoutput=1
\documentclass[
    ,final            % use final for the camera ready runs
  ]
  {aipproc}

\layoutstyle{6x9}

\pdfoutput=1

\newcommand{\etal}{{et~al.}~}

\def\arcsec{\hbox{$^{\prime\prime}$~}}

\begin{document}

\pdfoutput=1

\title{Tracing Star Formation in Cool Core Clusters with GALEX}

\classification{98.62.Ai; 95.85.Mt}
\keywords      {External galaxies and extragalactic objects: Origin, formation, evolution, age, and star formation; Astronomical Observations: Ultraviolet ($10-300$ nm)}

\author{Amalia K. Hicks}{
  address={Department of Physics \& Astronomy, Michigan State University, East Lansing, MI 48824-2320.}
}

\author{Richard F. Mushotzky}{
  address={Goddard Space Flight Center, Code 662, Greenbelt, MD, 20771.}, 
  altaddress={Department of Astronomy, University of Maryland, College Park, MD 20742-2421.} % additional visiting address
}

\author{Megan Donahue}{
  address={Department of Physics \& Astronomy, Michigan State University, East Lansing, MI 48824-2320.}
}

\begin{abstract}
We present recent results from a GALEX investigation of star formation in 16 cooling 
core clusters of galaxies, selected to span a broad range in both redshift and central cooling time. 
Initial results demonstrate clear UV excesses in most, but not all, brightest cluster galaxies in our 
sample. This UV excess is a direct indication of the presence of young massive stars and, 
therefore, recent star formation. We report on the physical extent of UV emission in these 
objects as well as their FUV-NUV colors, and compare GALEX inferred star formation rates to 
central cooling times, H$\alpha$ and IR luminosities for our sample. 
\end{abstract}

\maketitle

%%%%%%%%%%%%%%%%%%%%%%%%%%%%%%%%%%%%%%%%%%%%
%% MAINMATTER
%%%%%%%%%%%%%%%%%%%%%%%%%%%%%%%%%%%%%%%%%%%%

\section{INTRODUCTION}

Despite past evidence of star formation in ``cooling flow'' (hereafter referred to as 
CF) clusters \cite[e.g.,][]{mcnamara89,crawford99} the fact that star formation rate (SFR) estimates differed 
drastically from inferred X-ray cooling rates led to doubt that the two phenomena were 
related. However, recent UV investigations \cite{mittaz01,hicks05}, Spitzer data \cite{quillen08,odea08}, and precision 
optical photometry \cite{bildfell08} have definitively shown that CF clusters are the sites of star 
formation, and that there is an indisputable relationship between X-ray properties and 
SFRs. Here we confirm and quantify this connection, using GALEX 
observations of a sample of 16 CF clusters.

\section{SAMPLE AND OBSERVATIONS}

The Galaxy Evolution Explorer (GALEX) is a space telescope with both imaging 
and spectroscopic capabilities in two ultraviolet wavebands, Far UV (FUV) $1350-1780$~\AA ~and Near UV (NUV) $1770-2730$~\AA ~\cite{martin05}. Our GALEX targets consist of 16 clusters of 
galaxies that exhibit strong evidence of central cooling.  These objects were chosen to 
cover a wide range in redshift ($0.02 < z < 0.45$) and central (R $=20$ kpc) cooling time 
($0.5 < \rm{t}_{\rm{cool}} < 2.5$ Gyr). Table 1 lists the objects in our sample, their redshifts, and 
GALEX exposure times. All of our targets were easily detected in both GALEX 
wavebands, with an average SNR of 40 (21) in the NUV (FUV), and minimum SNRs 
of $\sim6$ in each band.

%Some url test \url{http://www.world.universe}.
%\newpage

%%%%%%%%%%%%%%%%%%%%%%%%%%%%%%%%%%%%%%%%%%%%
%% SAMPLE TABLE
%%
%% Shows the use of \tablehead and \tablenote
%% macros
%%%%%%%%%%%%%%%%%%%%%%%%%%%%%%%%%%%%%%%%%%%%

\begin{table}
\begin{tabular}{lccclcc}
\hline
  \tablehead{1}{l}{b}{Cluster}
  & \tablehead{1}{c}{b}{$z$}
  & \tablehead{1}{c}{b}{Exposure\\(NUV/FUV) [s]}
  & \tablehead{1}{c}{b}{}   
  & \tablehead{1}{l}{b}{Cluster\\(cont.)}
  & \tablehead{1}{c}{b}{$z$\\(cont.)}
  & \tablehead{1}{c}{b}{Exposure\\(cont.)}   \\
\hline
Abell 85 & 0.0557 & 2494 / 2494 & & Hydra A   & 0.0549 & 2230 / 2230\\
Abell 1204 & 0.1706 & 3738 / 3738 & & MKW3s   & 0.0453 & 2271 / 2271\\
Abell 2029 & 0.0779 & 1517 / 1517 & & MKW4   & 0.0196& 2194 / 2194\\
Abell 2052 & 0.0345 & 2863 / 2863 & & MS0839.8+2938   & 0.1980 & 4729 / 4728\\
Abell 2142 & 0.0904 & 1556 / 1556 & & MS1358.4+6245  & 0.3272 & 5614 / 5614\\
Abell 2597 & 0.0830 & 2111 / 2111 & & MS1455.0+2232  & 0.2578 & 3385 / 3384\\
Abell 3112 & 0.0761 & 4873 / 2618 & & RXJ1347.5-1145  & 0.4500 & 9120 / 9119\\
Hercules A & 0.1540 & 3870 / 3870 & & ZwCl 3146  & 0.2906 & 3127 / 3127\\
\hline
\end{tabular}
%\caption{}
\label{tab:a}
\end{table}

\section{IMAGING AND PHOTOMETRY}

Surface brightness profiles in both wavebands were constructed for our targets in $5\arcsec$ 
bins (the approximate size of the larger PSF). Some of these profiles indicate UV 
emission at greater radii than had been previously detected. The SB profiles were then 
used to create radial color profiles. These tend to indicate pure star formation at small 
radii, then become progressively redder until, at large radii, colors are consistent with 
those of non-star forming ellipticals \cite{gildepaz07}. An example of each is shown in Figure 1.  

Photometry was performed in $7\arcsec$ radius apertures centered on each BCG. This 
aperture size was chosen for compatibility with readily available 2MASS photometry. Using multiple archival GALEX observations, a control sample 
of 24 quiescent cluster ellipticals was obtained. These objects were used to construct a 
calibration relationship between UV and J band emission; effectively predicting the 
amount of UV light ``expected'' from a given old stellar popuation. This relationship 
was used to quantify the amount of ``excess'' UV light emitted by our CF sample 
(Figure 2). The majority of our sample exhibits clear UV excesses, indicating recent 
star formation.

%%%%%%%%%%%%%%%%%%%%%%%%%%%%%%%%%%%%%%%%%%%%
%% Sample figure:
%%
%% The option [height=...] scales the picture to the given height,
%% without it it would be printed at its nominal size
%%%%%%%%%%%%%%%%%%%%%%%%%%%%%%%%%%%%%%%%%%%%

\begin{figure}[h!]
  \includegraphics[height=.22\textheight]{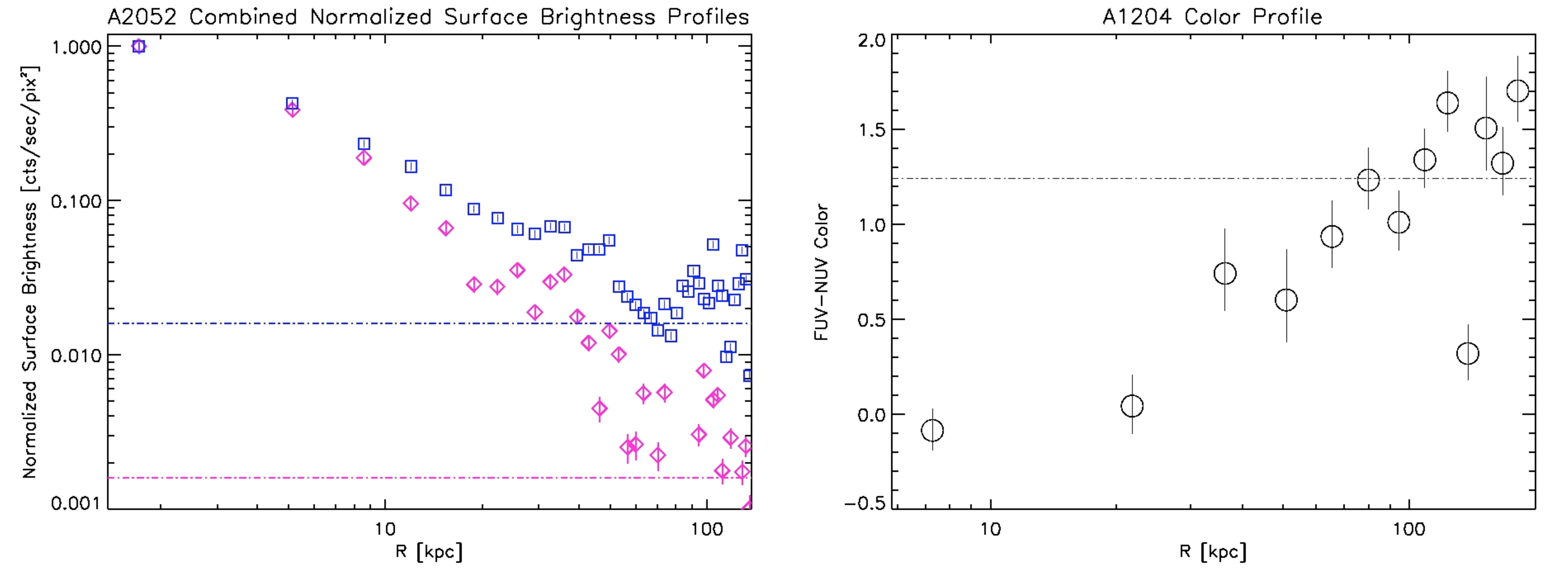}
  \caption{{\it{Left (a)}}: UV surface brightness profiles for Abell 2052, normalized to a value of 1 in the 
central bin. The NUV profile is plotted with square symbols and FUV with diamonds.  Horizontal lines 
indicate average NUV (top) and FUV (bottom) background levels. Note that UV emission (i.e., star 
formation) is detectable out to unprecedented radii with GALEX. {\it{Right (b)}}: UV color profile (FUV- 
NUV) for Abell 1204, clearly showing bluer emission in the center. The horizontal line designates the 
color of the typical GALEX background. Star forming galaxies tend to have GALEX UV colors $\sim0.4$, 
while the majority ($82\%$) of passively evolving elliptical galaxies have UV colors of $>0.9$ \cite{gildepaz07}.}
\end{figure}

%%%%%%%%%%%%%%%%%%%%%%%%%%%%%%%%%%%%%%%%%%%%
%% Sample figure:
%%
%% The option [height=...] scales the picture to the given height,
%% without it it would be printed at its nominal size
%%%%%%%%%%%%%%%%%%%%%%%%%%%%%%%%%%%%%%%%%%%%

\begin{figure}[h!]
  \includegraphics[height=.21\textheight]{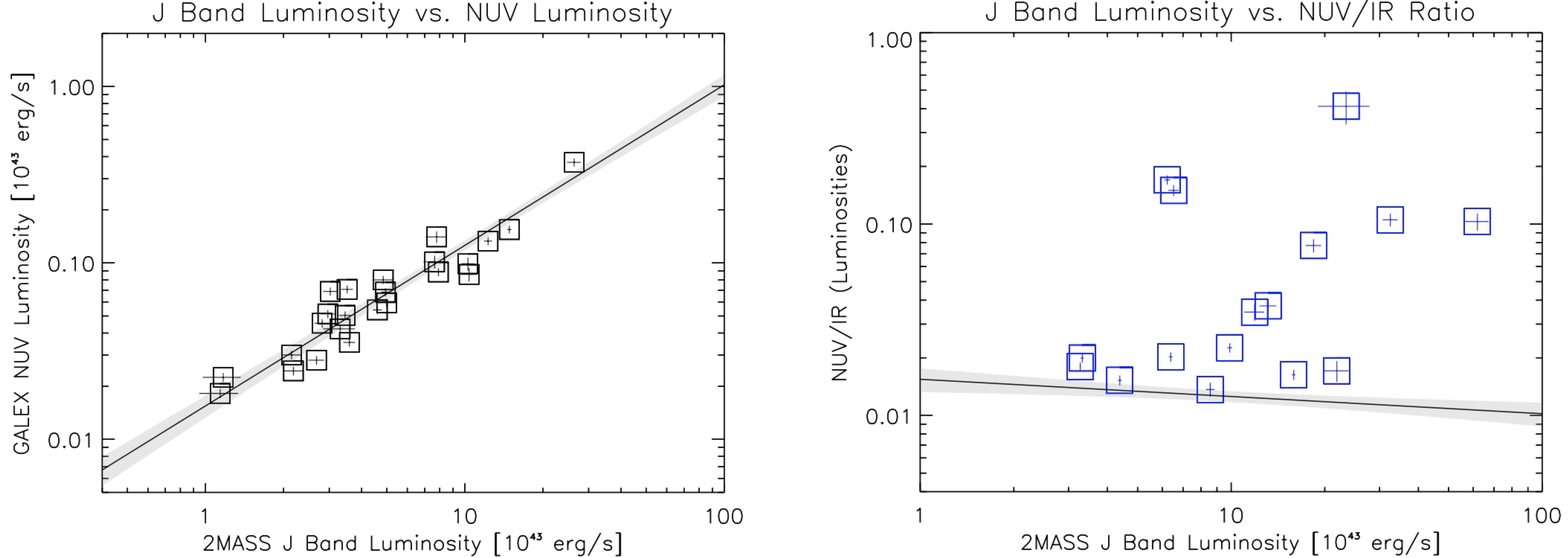}
  \caption{{\it{Left (a)}}: NUV/J band calibration relationship obtained from 24 cluster 
ellipticals. {\it{Right (b)}}: NUV/J vs. J band luminosity for our CF sample (squares). The line 
indicates the expected relationship for passively evolving ellipticals from our correlation; shaded regions designate $1\sigma$ errors on the fit.}
\end{figure}

\section{MULTIWAVELENGTH COMPARISONS}

For comparative purposes only, we translated our UV excesses into star formation 
rates using a Starburst99 model for continuous star formation over 20 Myr \cite{leitherer99}. This 
timescale was chosen to grossly approximate episodic cooling timescales (during 
which the system undergoes feedback processes with alternating heating and cooling 
cycles). These SFRs were then compared to cluster properties from the literature.   

H$\alpha$ measurements for our sample were taken from \cite{crawford99,donahue92,cavagnolo09}, and are shown vs. 
UV inferred SFRs in Figure 3. The consistency between UV and H$\alpha$ inferred star 
formation rates is remarkable, despite the many assumptions and unknowns that 
plague such comparisons. Infrared fluxes came from \cite{quillen08,edge01,egami06,donahue07}, and were converted 
to SFRs as in \cite{odea08}. This comparison plot is also shown in Figure 3.  

The X-ray properties of our sample come from \cite[][plus private comm.]{cavagnolo09}.  In Figure 4 we show UV 
inferred SFRs vs. both entropy and cooling time at 20 kpc from the cluster center. The 
correlation between UV SFR and cooling time proves conclusively that the star 
formation in these objects is directly related to cooling gas in the cluster cores. 

%USE KENS THESIS???  IS THERE ROOM???

%%%%%%%%%%%%%%%%%%%%%%%%%%%%%%%%%%%%%%%%%%%%
%% Sample figure:
%%
%% The option [height=...] scales the picture to the given height,
%% without it it would be printed at its nominal size
%%%%%%%%%%%%%%%%%%%%%%%%%%%%%%%%%%%%%%%%%%%%

\begin{figure}[h!]
  \includegraphics[height=.21\textheight]{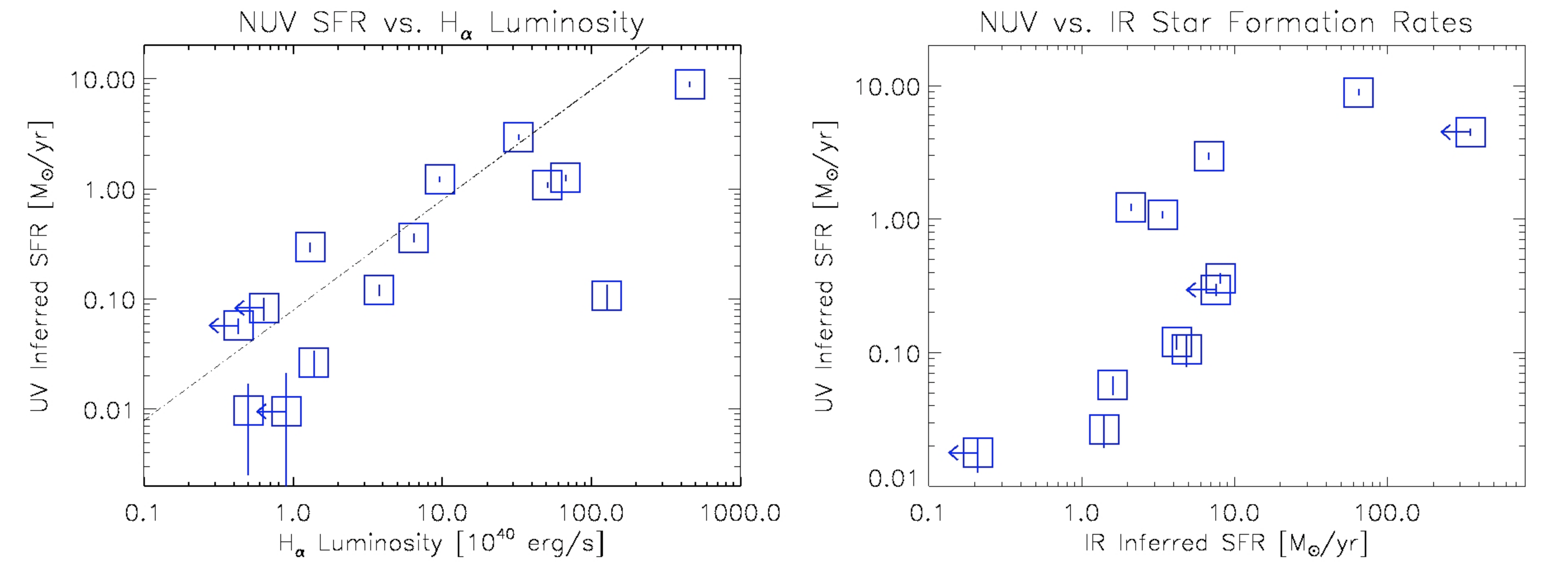}
  \caption{{\it{Left (a)}}: UV inferred SFR vs. H$\alpha$ luminosity for a subset of our targets. The line indicates the L$_{\rm{H}\alpha}$-SFR relationship of \cite{kennicutt98}; {\it{Right (b)}}: UV vs. IR inferred SFRs. A factor of $\sim10$ discrepancy may suggest that star formation is highly obscured in our targets, however, this is unconfirmable at present due to the many assumptions inherent in SFR conversions in both wavebands.}
\end{figure}

%%%%%%%%%%%%%%%%%%%%%%%%%%%%%%%%%%%%%%%%%%%%
%% Sample figure:
%%
%% The option [height=...] scales the picture to the given height,
%% without it it would be printed at its nominal size
%%%%%%%%%%%%%%%%%%%%%%%%%%%%%%%%%%%%%%%%%%%%

\begin{figure}[h!]
  \includegraphics[height=.21\textheight]{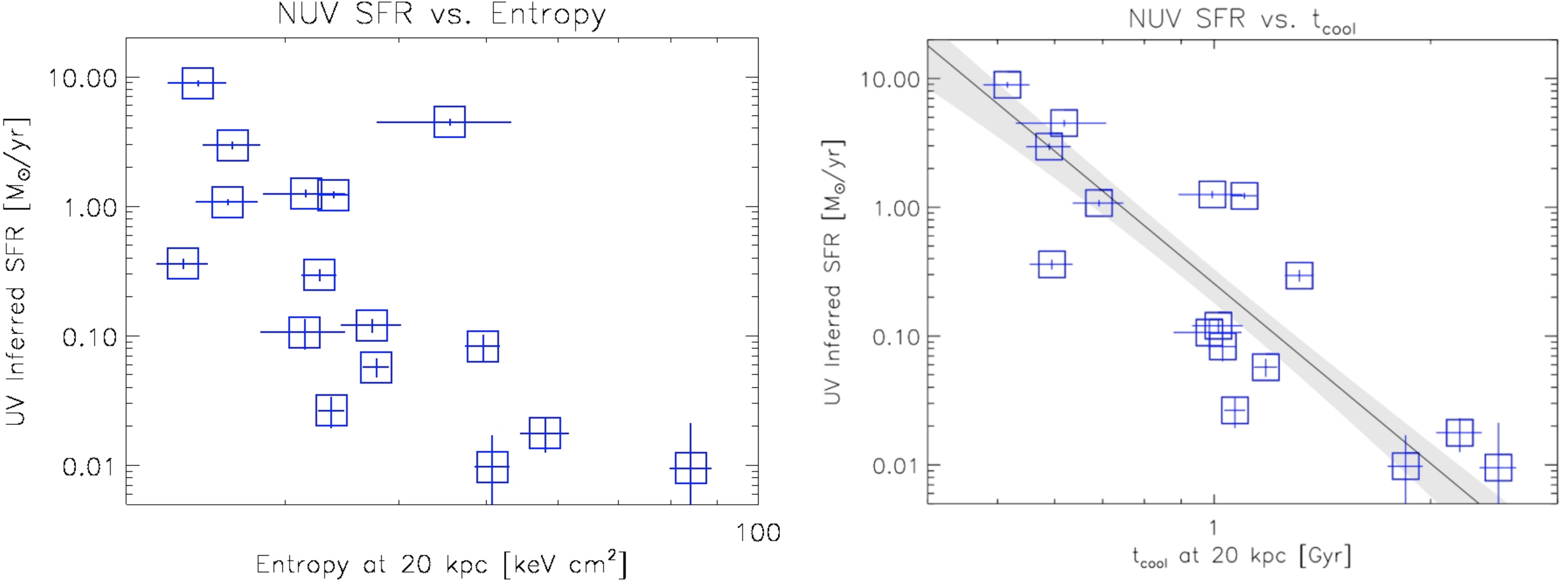}
  \caption{{\it{Left (a)}}: UV inferred SFR vs. central gas entropy (R $=20$ kpc), showing a tendency for 
more star formation to occur in lower entropy objects. The outlying point is our highest-$z$ cluster. {\it{Right (b)}}: UV inferred SFR vs. central cooling time (R $=20$ kpc). A line shows the best BCES regress bisector fit, with a slope of $-4.7\pm0.5$. The shaded region designates $1\sigma$ errors on the fit.}
\end{figure}

\section{SUMMARY}

GALEX easily detects star formation in cluster BCGs out to $z\ge 0.45$ and to 
unprecedented radii. In most of the CF clusters studied, we find significant UV 
luminosity excesses and colors in the central galaxies that together suggest recent 
and/or current star formation. This finding is corroborated by H$\alpha$ and IR observations. 
A correlation between UV excess and central cooling time confirms that this star 
formation is directly and incontrovertibly related to the cooling gas.

%%%%%%%%%%%%%%%%%%%%%%%%%%%%%%%%%%%%%%%%%%%%%%%%
%% BACKMATTER
%%%%%%%%%%%%%%%%%%%%%%%%%%%%%%%%%%%%%%%%%%%%%%%%

\begin{theacknowledgments}
Support for this work was provided by NASA through GALEX award NNX07AJ38G and LTSA grant NNG05GD82G.
\end{theacknowledgments}

\end{document}